\documentclass[12pt]{article}
\usepackage{jheppub}
\usepackage{epsfig}
\usepackage{amssymb}
\usepackage{verbatim}
\usepackage{amsmath}
\usepackage{dsfont}
\usepackage{slashed}
\usepackage[dvipsnames]{xcolor}
\usepackage{epstopdf}
\usepackage{graphicx}
\definecolor{light-gray}{gray}{0.50}

\def\D{\Delta}

\newif\ifdraft
\drafttrue
\newif\ifpreprint
\preprinttrue

\def\ba#1\ea{\begin{align*}#1\end{align*}}

\def\spa#1.#2{\left\langle#1\,#2\right\rangle}
\def\spb#1.#2{\left[#1\,#2\right]}

\def\e{\epsilon}

\def\hp{\hat{p}}

\newcommand{\eq}{\begin{equation}}
\newcommand{\eqe}{\end{equation}}
\newcommand{\eqa}{\begin{eqnarray}}
\newcommand{\eqae}{\end{eqnarray}}

\def\det{{\rm Det}}

\newbox\charbox
\newbox\slabox
\def\s#1{{      
        \setbox\charbox=\hbox{$#1$}
        \setbox\slabox=\hbox{$/$}
        \dimen\charbox=\ht\slabox
        \advance\dimen\charbox by -\dp\slabox
        \advance\dimen\charbox by -\ht\charbox
        \advance\dimen\charbox by \dp\charbox
        \divide\dimen\charbox by 2
        \raise-\dimen\charbox\hbox to \wd\charbox{\hss/\hss}
        \llap{$#1$}
}}

\title{
S-matrix singularities and CFT correlation functions} 

\author{
Carlos Cardona$^{a}$, Yu-tin Huang$^{a,b}$}

\affiliation{$^a$ Physics Division, National Center for Theoretical Sciences, National Tsing-Hua University,
No.101, Section 2, Kuang-Fu Road, Hsinchu, Taiwan}
\affiliation{$^b$ Department of Physics and Astronomy, National Taiwan University, Taipei 10617, Taiwan}

\abstract{
In this note, we explore the correspondence between four-dimensional flat space S-matrix and two-dimensional CFT proposed by Pasterski et al. We demonstrate that the factorization singularities of an $n$-point cubic diagram reproduces the AdS Witten diagrams if mass conservation is imposed at each vertex. Such configuration arises naturally if we consider the $4$-dimensional S-matrix as a compactified massless $5$-dimensional theory. This identification allows us to rewrite the massless S-matrix in the CHY formulation, where the factorization singularities are re-interpreted as factorization limits of a Riemann sphere. In this light, the map is recast into a form of $2d/2d$ correspondence.   
}
\begin{document}

\begin{flushright}
\vspace{10pt} \hfill{NCTS-TH/1703} \vspace{20mm}
\end{flushright}

\maketitle

\section{Introduction}  
The isomorphism between the Lorentz group in four dimensions and the Mobius group of conformal transformations in two dimensions, have been used in several theoretical approaches for the computation of scattering amplitudes in the past. It is indeed the corner stone of the realization of  Penrose's Twistor Space \cite{Penrose:1967wn} and the impressive subsequent development of scattering amplitudes of massless particles in Twistor space, started by Witten more than a decade ago \cite{Witten:2003nn}.  A generalisation of this isomorphism is the Embedding Formalism for the $d$-dimensional conformal group, built upon the work of Dirac \cite{Dirac:1936fq} and which has been particularly useful in the context of the AdS/CFT correspondence 

Recently, there has been a growing effort in writing the dynamics of four-dimensional Minkowski space in terms of observables in two-dimensional conformal field theory, greatly  motivated by the renewed interest on the asymptotic BMS symmetries in gravitational theories
\cite{Bondi:1962px, Sachs:1962zza} (for a more recent discussion see \cite{Haco:2017ekf}), based on the observation that the Lorentz group in four-dimension acts as the Mobius group in two dimensions over the null-infinity boundary of the three-dimensional space, recently baptized as the celestial sphere $\mathcal{CS}$ \cite{Cheung:2016iub}. 
From this point of view, it is expected that scattering amplitudes in four-dimensions can be recasted in terms of some sort of correlator in a certain two-dimensional conformal field theory. It has indeed been shown that soft theorems can be rewritten as Ward identities in a two-dimensional conformal field theory \cite{Lipstein:2015rxa, Strominger:2013lka} and hence, they should be somehow related to two-dimensional current algebras 
\cite{ He:2015zea, Cardona:2015woa}. The conserved currents and the stress-tensor of the corresponding  two-dimensional field theory has been discussed in \cite{Cheung:2016iub, Kapec:2016jld}.


More recently, Pasterski, Shao and Strominger considered S-matrix for conformal primary wave functions. The wave functions are constructed by convoluting plane waves with bulk-to-boundary propagators in $AdS_3$, and thus transforms covariantly under the Mobius group \cite{Pasterski:2016qvg}.
Using this transformation allows one to transform Lorentz invariant (Little group covariant) scattering amplitudes into Mobius covariant quantities, which can be considered as the correlation functions of some 2d CFT. The precise map is:
 \begin{align} &\mathcal{\tilde A}(\Delta_i,w_i,\bar w_i) 
\equiv \prod_{i=1}^n\left(\int d^2z_i\frac{dy_i}{y_i^3}G_\Delta (y_i, z_i,\bar z_i; w,\bar w) \right)\mathcal{A}(m_j\hat{p}_j)\,,
\end{align} 
where we have $n$ copies of integration over AdS$_3$ coordinates, and $G_\Delta$ are the bulk to boundary propagators. The AdS$_3$ coordinates are imbedded in the four-dimensional momenta satisfying the massive on-shell constraint $p^2=-m^2$. Explicit results were obtained for the three-point function of $\phi^3$ theory, which is fixed by symmetries.

In this note we intend to explore this relation further by studying the convolution of factorization singularities of flat space scattering amplitudes. We consider the factorization singularities of massive scattering amplitudes that admit a cubic diagram expansion. By ensuring that the mass is conserved at each vertex, we show that the factorization singularity for massive poles, enforces that the $n$-point kinematics can be mapped to a configuration of a contact AdS Witten diagram. Once dressed with bulk to boundary propagators and integrate over the whole AdS space, one reproduce an $n$-point correlation function, i.e.:
\eq 
\vcenter{\hbox{\includegraphics[scale=0.5]{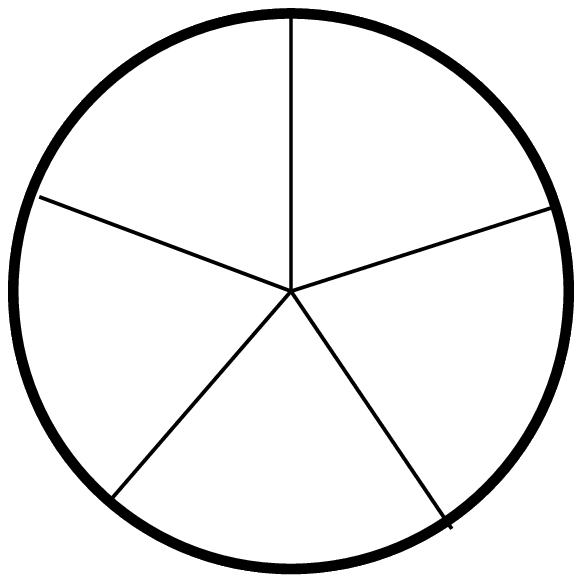}}}\;\;=\;\; \prod_{i=1}^n\left(\int d^2z_i\frac{dy_i}{y_i^3}G_\Delta (y_i, z_i,\bar z_i; w,\bar w) \right)\vcenter{\hbox{\includegraphics[scale=0.5]{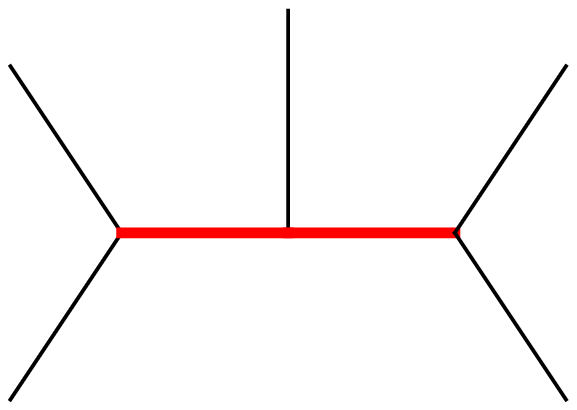}}}\,,
\eqe
where on the cubic diagram on the RHS we identify the pole singularities. Since all tree-level scalar Witten diagrams can be re-expressed in terms of finite sums of contact diagrams~\cite{DHoker:1998ecp, DHoker:1999mqo, Penedones:2010ue}, with the scaling dimensions of external operators being shifted, this implies that the tree-level S-matrix singularities reproduces all tree-level Witten diagrams.

Note that mass conservation can be interpreted as momentum conservation in higher dimensions, where the propagators are massless. The embedding of the previous construction in terms of a massless theory allows us to introduce the Cachazo-He-Yuan (CHY) representation for the propagator singularities. In particular, as discussed in~\cite{Cachazo:2013iea}, by parameterising the moduli space of $n$-punctures by a single parameter $\tau$ such that 
\eq
\sigma_i=\frac{v_i}{\tau}\;\;i\in L,\quad \sigma_i=\tau u_i\;\; i\in R\,.
\eqe
where $\sigma_i$ are the coordinates of the punctures and we've separated it into a left and right set, the parameter $\tau$ then encodes the pinch limit of the Reimann sphere. More over, in such parameterization, one can identify that one of the scattering equation constraints becomes 
\eq
\delta(\tau^2F-p^2_I)
\eqe
where $p_I$ would be the associated momenta of the internal particle and $F$ is some $\tau$ independent polynomial. Thus by inserting a factor of $\delta(\tau)$ in the CHY integrand, we reproduce the factorization constraint. This combined with the fact that there the CHY formula can be naturally computed from a worldsheet chiral string theory~\cite{Mason:2013sva, Huang:2016bdd}, give an interesting $2d/2d$ correspondence: 
\eq
2d\;\;CFT\; \leftrightarrow  \prod_{i=1}^n\left(\int d^2z_i\frac{dy_i}{y_i^3}G_\Delta (y_i, z_i,\bar z_i; w,\bar w) \right)\; \int  d^{2n}\sigma_i \;{\rm CHY} \;\left(\prod_j \tau_j\delta(\tau_j)\right)\,,
\eqe 
where $\tau_i$ are moduli for the degenerate limits of punctures on the Riemann sphere. 

The remaining of this paper is organized as follows: in section 2 we quickly review the Pasterski-Shao-Strominger (PSS) proposal for the transformation of scattering amplitudes in flat space into correlation functions of conformal field theory at co-dimension two, then in section 3 we move to the study of the transform at the factorization singularities of the S-matrix. Finally, at section 4 we propose a duality relation between correlation functions in two-dimensional chiral string theory in the CHY representation of the S-matrix and correlation functions from the PSS transformation.

\textit{During the preparation of this draft, the authors were aware of the coming work by Dhritiman Nandan, Anastasia Volovich, Congkao Wen, and Michael Zlotnikov, which has some overlap. }
\section{The PSS proposal}
The fact that $d$-dimensional conformal symmetry can be linearly realised as a $d+2$-dimensional Lorentz symmetry has a long history of applications that dates back to Dirac (see \cite{Siegel:2012di} for references and review). It is then natural to ask whether or not observables on both sides of the relation can also be mapped. For CFTs, the natural physical observables are correlation functions, which transform covariantly under conformal transformations. On the $d+2$-dimensional side, the natural Lorentz covariant physical observable is the S-matrix, which upon Lorentz transformation generates a little transformation. The next task is then to find a map between the variables on the two sides.

Recently Pasterski, Shao and Strominger (PSS) \cite{Pasterski:2016qvg} presented a proposal for such a map, where as a first step, the four-dimensional massive momenta $\mathbb{R}^{1,3}$ is mapped into the coordinates of hyperbolic space $H_3$. Taking the metric of $H_3$ as,
\begin{align}\label{ads3metric}
ds_{H_3}^2 = {dy^2 +dzd\bar z\over y^2}\,,
\end{align}
the three-dimensional hyperbolic space is mapped to the four-dimensional momenta in a SL(2,C) covariant form as:
\eq
\hat{p}_{a\dot{a}}(y,z)\equiv \hat{p}^ {\mu}(y,z)\sigma_{\mu}=\frac{i}{y}\begin{bmatrix}1 & \bar{z}\\
z & y^2+|z|^2
\end{bmatrix}\label{momentum}
\eqe
such that $\hat p^2=\det[\hat{p}_{a\dot{a}}]=-1$. For a particle of mass $m$, it's momenta is given as $p=m\hat{p}$.

It was conjectured that the correlation function of some 2D CFT can be related to the S-matrix in four-dimensions, where the external states are conformal primary wave-functions, defined as:
\eq\label{integral}
\phi^\pm_{\Delta,m} (X^\mu; w, \bar w)= 
\int_0^\infty   {dy \over y^3} \int dzd\bar z  \, 
G_\Delta (y ,z,\bar z;w, \bar w)  \,
\exp\Big[   { \pm i m \,\hat p^\mu(y,z,\bar z)\, X_\mu}   \Big]
\eqe
where the $\pm$ in the exponent indicates the incoming and outgoing states and $G_\Delta(y,z,\bar z; w,\bar w)$ be the scalar bulk-to-boundary propagator in $H_3$ of conformal dimension $\Delta$ \cite{Witten:1998qj}, 
\begin{align}
G_\Delta (y, z,\bar z; w,\bar w) =  \left({ y\over y^2  +  |z-w|^2}\right)^\Delta\,.
\end{align}
Note that we can write the bulk to boundary propagator as 
\eq 
G_\Delta (y, z; w)=\left(
-i \begin{bmatrix}
w & 1
\end{bmatrix}^a
\hat{p}(y,-z)_{a,\dot{a}}
\begin{bmatrix}
\bar{w}\\1
\end{bmatrix}^{\dot{a}}\right)^{-\Delta}\,,
\eqe
which manifests it's covariant property under the $SL(2,\mathbb{C})$ transformation $w' = (aw+b)/(cw+d)$,\footnote{Here and in what follows we have only used the dependence on variable $z$, both for simplify the notation and because we are not taken $\bar{z}$ to be an independent variable, but instead as the conjugate of $z$. },
\begin{align}\label{covariance}
&G_\Delta(y',z',\bar z'; w' ,\bar w') = |cw+d|^{2\Delta} G_\Delta(y,z,\bar z; w,\bar w)\,.
\end{align}

The correlation function $\mathcal{\tilde A}_{\Delta_1,\cdots,\Delta_n}(w_i,\bar w_i) $ is then related to the flat space S-matrix as defined as $\mathcal{A}(m\hat{p})$ by:
 \begin{align}\label{transform}
 &\mathcal{\tilde A}(\Delta_i,w_i,\bar w_i) 
\equiv \int d^4X\,\prod_{i=1}^n \phi^\pm_{\Delta_i,m_i} (X^\mu; w_i, \bar w_i)\mathcal{A}(m_j\hat{p}_j).
\end{align}
The four-dimensional integral $d^4X$ simply produces the momentum conservation delta function. The plausibility of eq.(\ref{transform}) stems from the two sides sharing the same symmetry, as verified in~\cite{Pasterski:2016qvg}, for the case of $\phi^2\varphi$ interaction, where $m_{\varphi}\sim2m_{\phi}$.

\subsection{The three point contact term}
The explicit example shown in~\cite{Pasterski:2016qvg} is a three-point contact term, and the mass of one particle is near-extremal, i.e. it is near the sum of the other two. Here we present a brief review, since most of the details will be utilised for the $n$-point construction.

Taking the mass of the first particle $\varphi$ to be $2(1+\e) m$ and the masses of the other two particles be $m$ and evaluating the $X^\mu$-integral, we arrive at the following expression for the scalar three-point amplitude,
\begin{align}
\mathcal{ \tilde A}(w_i,\bar w_i)&=i(2\pi)^4 \lambda m^{-4}
\left( \prod_{i=1}^3 \int_0^\infty   {dy_i \over y_i^3} \int dz_id\bar z_i \, \right)\times\nonumber\\
&~~~\prod_{i=1}^3 G_{\Delta_i} (y_i , z_i,\bar z_i ;  w_i,\bar w_i)\,
\delta^{(4)}(-2(1+\e) \hat p_1 +\hat p_2+\hat p_3)\,,
\end{align}
where we used $-p_1+p_2+p_3=-2(1+\e) \hat p_1 +\hat p_2+\hat p_3$. In general, three of the four momentum conservation delta function solves one of the momenta in terms of others, while the remaining one simply enforces 
\eq \label{4condition}
p_n^2=(\sum_{i=1}^{n-1}p_i)^2=-m_n^2
\eqe

For the current case the integral of $(y_3, z_3)$ is straightforwardly localized, leaving a Jacobian factor of 
\eq
-\left(-2(1+\epsilon)\frac{y^2_1+|z_1|^2}{y_1}+\frac{y^2_2+|z_2|^2}{y_2}\right)^{-1}
\eqe
and we are left with the final delta function, which is proportional to the on-shell condition, 
\eqa
&&\delta\left(2(1{+}\epsilon)\frac{1}{y_1}{-}\frac{1}{y_2}{-}\frac{1}{y_3}\right)\nonumber\\
&=&\left({-}2(1{+}\epsilon)\frac{y^2_1{+}|z_1|^2}{y_1}{+}\frac{y^2_2{+}|z_2|^2}{y_2}\right)\delta(\hat{p}_1\cdot\hat p_2+(1{+}\e))\nonumber
\eqae
Note that the prefactor on the RHS exactly cancels the previous Jacobian factor. Thus one concludes that momentum conservation fixes the position of one leg:
\eq\label{Solution}
z_{n}=-y_n\left(\sum_{i}m_i\frac{z_i}{y_i}\right),\quad y_n=\frac{-\sum_{i}m_i\frac{y_i^2+|z_i|^2}{y_i}}{1+\left(\sum_{ij}\frac{m_im_jz_i\bar{z}_j}{y_iy_j}\right)}
\eqe
where the summation range is $i=1,\cdots,n{-}1$ and appropriate additional signs for each term depending on whether it is incoming or outgoing.

It is convenient to parameterize the remaining two AdS$_3$ points as
\eq
y_2=y_1+R\cos\theta,\quad z_2=z_1+R\sin\theta e^{i\phi}\,,
\eqe
for which the argument of the delta function becomes
\eq\label{localiza}
\delta(\hat p_1\cdot\hat p_2+(1{+}\e))=2y_1y_2\delta(2\epsilon y_1y_2-R^2)\,.
\eqe
Thus the final constraint forces $(y_2, z_2)\rightarrow (y_1, z_1)$, and through eq.(\ref{Solution}), so is $(y_3, z_3)$, arriving at a contact diagram in AdS$_3$. Note that if $\epsilon<0$, there will be no solution for the delta function since it is below the production threshold. 

The support of the delta function in equation \eqref{localiza}, at leading order in $\epsilon$, is given by,
\eq \label{RtoY}
R=\sqrt{2\epsilon}\,y_1=\tilde{R}\sqrt{\epsilon}\,,
\eqe
where $\tilde{R}=y_1\sqrt{2}$. Therefore, the integration measure translate into the coordinate $(\tilde{R},\theta,\phi)$ as,
\eq\label{Rmeasure}
\int d^2z_1\,dy_1\to\epsilon^{3/2}2\pi\int \tilde{R}^2 d\tilde{R}\,d\Omega\,.
\eqe
So more precisely, we are retaining the leading order in small $\epsilon$ expansion. In the following sections we are going to use heavily the condition \eqref{localiza}, therefore the final transformation should be understood as the leading piece in  $\epsilon$. More explicitly for the three-point function considered in this section we write,
 \vspace{5mm}
 \begin{align}\label{vertex}
\mathcal{ \tilde A}_3(w_i,\bar w_i)&\approx 2i\frac{(2\pi)^5}{m^4} \lambda
\epsilon^{1/2}\int_0^\infty   {dy_2 \over (y_2)^4} \int d^2z_2\times\nonumber\\
&~~~\int\tilde{R}^2 d\tilde{R}\,\int d\Omega\, \delta(\tilde{R}^2-2y_2^2)\prod_{i=1}^3 G_{\Delta_i} (y_i , z_i,\bar z_i ;  w_i,\bar w_i)\nonumber\\
&\sim \lambda\frac{\epsilon^{1/2}}{m^4} 
\int_0^\infty   {dy_2 \over y_2^3} \int d^2z_2\prod_{i=1}^3 G_{\Delta_i} (y_i , z_i,\bar z_i ;  w_i,\bar w_i)\,.
\end{align}
\section{Contact diagrams from factorization singularities}
As mentioned previously, an interesting questions one can immediately pose for such a correspondence is how consistency conditions are mapped. The S-matrix is subject to locality and unitarity constraints which translate into the presence of propagator singularities as well as consistent residues. Here we will explore the implication of factorization singularities on the $n$-copy AdS integral.
\subsection{A four-point example}
\begin{figure}
\begin{center}
\includegraphics[scale=0.6]{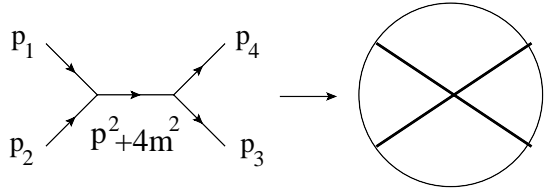}
\caption{The factorization singularity of $s$-channel diagram leads to a contact Witten diagram in AdS$_3$. }
\label{fig1}
\end{center}
\end{figure}

Let's assume that $(1, 2)$ are incoming and $(3,4)$ are outgoing with identical absolute mass. Mass conservation restricts the $s$-channel pole to be massive. Thus the complete kinematic constraint for $s$-channel factorization is given by: 
\eq\label{schannel}
\delta^{(4)}(\sum_i p_i)\delta((p_1+p_2)^2+4m^2)
\eqe   
In terms of $\hat{p}_i$ the argument of the delta functions can be written as 
\eq
\hat{p}_1\cdot\hat{p}_2+1,\quad{\rm or}\quad\hat{p}_3\cdot\hat{p}_4+1\,. 
\eqe
As seen from the previous section whenever the AdS coordinates appear under the constraint 
\eq\label{localization}
\hat{p}_i\cdot \hat{p}_j+1=0\,,
\eqe
points $i$ and $j$ will be forced to be coincidental. Indeed here if one performs a change of variable,
\eqa
y_{12}&=&R_{12}{\rm cos}\theta_{12},~~ z_{12}=R_{12}{\rm sin}\theta_{12}e^{i\phi_{12}}\nonumber\\
y_{34}&=&R_{34}{\rm cos}\theta_{34},~~ z_{34}=R_{34}{\rm sin}\theta_{34}e^{i\phi_{34}}\,.
\eqae
where $y_{ij}\equiv y_i-y_j$ and $z_{ij}\equiv z_i-z_j$, it is straightforward to see that this constraint simply forces points $1,2$ and $3,4$ to be coincidental respectively. Finally, three of the four momentum conservation delta function solves one of the momenta in terms of others, and the remaining constraint is \eqref{4condition}, which in this case leads to:
\eq 
\hp_1\cdot\hp_3+1=0\,.
\eqe
Therefore momentum conservation forces all points to be coincidental. In other words, we end up with a contact term in AdS$_3$ as indicated in fig.(\ref{fig1}). 

Let us put all the pieces together in this simple example, to illustrate how this will work in general. From equation \eqref{transform} for four-particles we have,
\begin{align}
\mathcal{ \tilde A}_4(w_i,\bar w_i)&=i(2\pi)^4 \frac{\lambda^2}{4m^{4}}\left( \prod_{i=1}^3 \int_0^\infty   {dy_i \over y_i^3} \int dz_id\bar z_i \, \right)\prod_{i=1}^4 G_{\Delta_i} (y_i , z_i,\bar z_i ;  w_i,\bar w_i)
\times\nonumber\\
&~~~\,
\delta(\hat p_1\cdot\hat p_2+1+\epsilon)\delta(\hat p_1\cdot\hat p_3+1)\,,
\end{align}
where should be understood that $p_4$ has been fixed by momentum conservation, the first delta function in the second line comes from the singularity pole \eqref{schannel}  whereas the second delta function comes from the remaining delta from momentum conservation \eqref{4condition}, namely $\delta(p_4^2+m^2)$.  From the explicit computation for the vertex \eqref{vertex}, we have also learned that upon integration, every localization delta function  $\delta(\hat p_i\cdot\hat p_j+1+\epsilon)$, give us a leading scaling $\epsilon^{1/2}$, so up to a numerical factor and at leading order in $\epsilon$ we find,
 \begin{align}
\mathcal{ \tilde A}_4(w_i,\bar w_i)\sim \lambda^2\frac{\epsilon^ {1/2}}{m^4} 
\int_0^\infty   {dy_1 \over y_1^3} \int d^2z_1\prod_{i=1}^4 G_{\Delta_i} (y_i , z_i,\bar z_i ;  w_i,\bar w_i)\,.
\end{align}
This is the contact four-point Witten diagram in $AdS_3$.

Note that even though we've only reproduced contact Witten diagrams, factorization Witten diagrams can be represented in a similar fashion. It was shown~\cite{DHoker:1998ecp, DHoker:1999mqo, Penedones:2010ue} that the latter can be rewritten as a finite series expansion in terms of contact quartic Witten diagrams,
\begin{align}\label{cont_exp}
 \sum_{k=1}^{\frac{2\D_0-\D}{2}}\frac{a_k}{|w_{13}|^{k}}&\int dp_1\,G_{k} (p_1 ;  w_1,\bar w_1)G_{k} (p_1 ;  w_3,\bar w_3)\times\nonumber\\
& G_{\D_0} (p_1 ;  w_2,\bar w_2)G_{\D_0} (p_1 ;  w_4,\bar w_4)\,,\nonumber\\
a_k=&\frac{(\D_0)_{-k}^2}{4\left(\frac{2\D_0-\D}{2}\right)_{1-k}\left(\frac{2\D_0+\D-d}{2}\right)_{1-k}}\,.
\end{align}
where $\D_0$ denotes the conformal dimension of the external fields, which in this note have been taken to be identical, and $\D$ denotes the conformal dimension of the exchanged field. Taking $w_1=0,w_3=1$, we find that factorization Witten diagrams can be identified as the residue of 
\eq
\mathcal{A}(\hat{p})=\sum_{k=1}^{\frac{2\D_0-\D}{2}}\frac{a_k}{(\hat{p}_1-\hat{p}_4)}\,.
\eqe
Note that the coefficients $a_k$ are all positive definite which is consistent from the view point of S-matrix unitarity.

Finally, since the constraint localizes all AdS points, the only possible Lorentz invariant is $\hat{p}_i\cdot\hat{p}_j=\hat{p}^2_i=-m_i^2$, i.e. the residue degenerates to a number. In other words, exchanging different higher spin states amounts to a trivial normalization constant for the contact AdS$_3$ diagram.\footnote{This can also be understood from the four-point kinematics $s$ and $t$. When $s=4m^2$,  
\eq \nonumber
t=\frac{s-4m^2}{2}(1-{\rm cos}(\theta))=0\,,
\eqe 
and thus the only possible residue is a number. }

\subsection{The $n$-point generalization}
We are now ready to make the full fledged proposal. We will consider all cubic graphs whose mass is conserved. The constraint arising from all massive propagators going on-shell, along with all momentum conservation, will reduce the $n$ copies of AdS integrals to a single copy, thus corresponding to a contact Witten diagram.

\begin{figure}
\begin{center}
\includegraphics[scale=0.5]{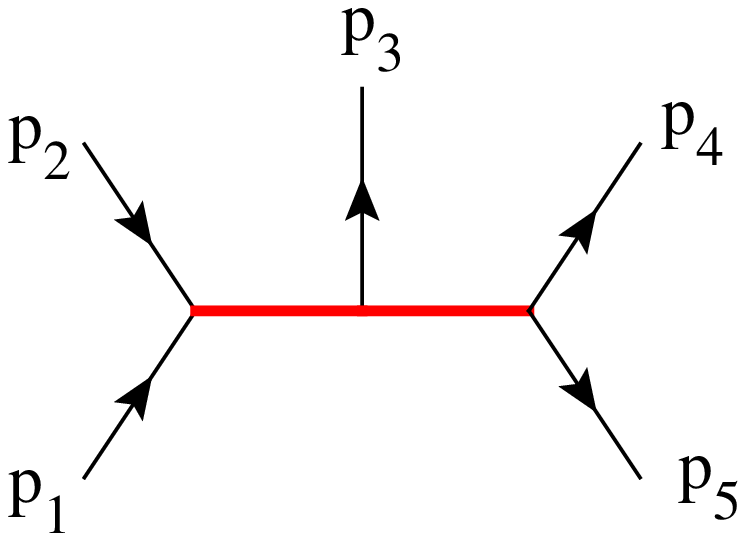}
\caption{}
\label{fig2.5}
\end{center}
\end{figure}

Consider the five-point cubic diagram represented in fig.(\ref{fig2.5}), where the arrows represent the moment flow for $p_1+p_2=p_3+p_4+p_5$.  The associated pole singularities are, 
$$\delta\left((p_1+p_2)^2+4m_{in}^2\right)\delta\left((p_3+p_4)^2+4m_{out}^2\right)$$ where $m_1=m_2=m_{in}$ and $m_3=m_4=m_5=m_{out}$. This  localizes  $\hat{p}_1\to \hat{p}_2$ and $\hat{p}_3\to \hat{p}_4 $.  Once again, momentum conservation can be used to solve for $\hat{p}_5$ and as in previous sections, the remaining on-shell condition becomes,
\eqa 
p_5^2&=&-m_{out}^2=[(p_1+p_2)-(p_3+p_4)]^2=4(-m_{in}^2-m_{out}^2-2m_{out}m_{in}\hat{p}_1\cdot\hat{p}_3)\,.
\eqae 
Mass conservation ensures that $m_{in}=3m_{out}/2$, and one again finds,
$$\hat{p}_1\cdot\hat{p}_3+1=0$$
which localizes $\hp_1\to \hp_3$, leading us to a single contact term in $AdS_3$.
\vspace{5mm}

{ \it $n$-point ladder diagrams}
\begin{figure}
\begin{center}
\includegraphics[scale=0.7]{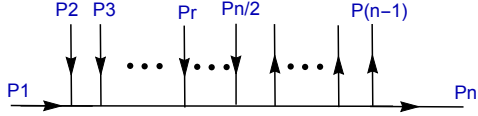}
\caption{{\it Half Ladder Feynman diagram }
\label{fig3}}
\end{center}
\end{figure}

The above analysis can be straightforwardly generalized to the scattering of $n-$particles for the half-ladder diagram shown at fig.(\ref{fig3}). Let us start considering an even number of particles such as, half of them are incoming with identical mass $m_{in}$ and the rest are out-going with also the same mass $m_{out}$, i.e $m_{in}=m_{out}$. Let the incoming particles to be adjacent. As before, mass conservation is imposed on each vertex. Momentum conservation allow us to fix one momentum, namely $p_n$. Following the order of the particle labels as in figure 3, the analysis is quite simple. The delta function singularity coming from the first propagator  $\delta\left((p_1+p_2)^2+4m_o^2\right)$, forces $\hp_1\to\hp_2$ as we have shown before. Then the singularity from the next vertex, namely,
\eqa 
\delta\left((p_1+p_2+p_3)^2+9m_o^2\right)=\delta\left((2p_1+p_3)^2+9m_o^2\right)\nonumber
\eqae
lead us to $\delta(\hp_1\cdot\hp_3+1)$, or $\hp_1\to\hp_3$. Continuing in this order, the singularity corresponding to the propagator at the right hand side of the vertex attached to $p_r$ with $r<n/2$  is given by,
\eq\delta\left(\left(\sum_{j=1}^rp_r\right)^2+\left(\sum_{j=1}^rm_o\right)^2\right)\
=\delta\left(\left({\over}(r-1)p_1+p_r\right)^2+(r\,m_o)^2\right)\,,
\eqe
which once again lead us to $\hp_1\to\hp_r$. We keep going until we reach the last propagator at the final vertex
and we use it to localize the leg $p_{n-2}$ along with the previous localized $n-3$ particles at the right of it \footnote{We can as well keep going until reach 
half of the diagram at leg $n/2$ and then perform the same analysis from right-to left until we reach the leg corresponding to particle $\frac{n}{2}+2$, in any case, we will arrive to the same conclusion}.  
Finally, the condition \eqref{4condition} plus mass conservation, lead us to
\eq 
p_n^2=-m_o^2=\left(\sum_{j=1}^{n/2}p_j-\sum_{j=n/2+1}^{n-1}p_j\right)^2=
\left(\frac{n}{2}p_1-\left(\frac{n}{2}-2\right)p_1-p_{n-1}\right)^2\,,
\eqe 
or equivalently 
\eq 
\hp_1\cdot\hp_{n-1}+1=0\,,
\eqe  
which implies $\hp_{n-1}\to \hp_1$ and therefore we end up again with a single contact term in $AdS_3$.

Now considering an odd-number of particles, with $(n-1)/2$ incoming and $(n+1)/2$ outgoing. Mass conservation tell us now 
\eq \label{massn}
m_{in}=\frac{n+1}{n-1}m_{out}
\eqe 
we perform a similar procedure as before, starting from left-to-right until we reach the leg $(n-1)/2$  and then the other way around until we reach the leg ${n+1\over 2}+1$. So finally the condition \eqref{4condition}, lead us to
\eq
p_n^2=\left(\sum_{j=1}^{(n-1)/2}p_j-\sum_{j=(n+1)/2}^{n-1}p_j\right)^2
=\left({(n-1)\over2}\right)^2(p_1-p_{n-1})^2\,,
\eqe
which after some algebra is equivalent to
\eq 
\hp_1\cdot\hp_{n-1}+1=0\,,
\eqe 
where we have used mass conservation condition \eqref{massn}.   Let us recall that on the  final transform we are retaining the leading order in an  $\epsilon$ expansion, and as we saw above, every delta function forcing a localization will contribute to a $\epsilon^{1/2}$ scaling, therefore, the Witten contact diagram coming from the $n-$particle half-ladder appear at order $\epsilon^{(n-3)/2}$,  coming from  $(n-3)$ propagators.\footnote{It is worth to recall that we are using $\epsilon$ to slightly get off from the singularity since on the singularity the transformation actually vanishes, so the role of the $\epsilon$ is to keep track of the order of the given zero.}
\vspace{5mm}

{ \it Benzene diagram}
\begin{figure}
\begin{center}
\includegraphics[scale=0.45]{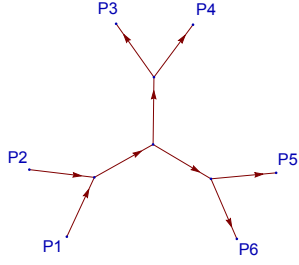}
\caption{Benzene-type diagram}
\label{}
\end{center}
\end{figure}

Now we consider the Benzene-type diagram represented in figure 4 for six-particles scattering with $p_1+p_2=p_3+p_4+p_5+p_6$. Looking at the following pole singularity, 
$$\prod_{j=\{1,3,5\}}\delta\left((p_j+p_{j+1})^2+4m_j^2\right)$$ where $m_1=m_2=m_{in}$ and $m_3=m_4=m_5=m_6=m_{out}$.  It localizes  $\hp_j\to \hp_{j+1}$ for $j=\{1,3,5\}$. Momentum conservation can be used to solve for $p_6$ and the additional condition \eqref{4condition} lead us to,
\eq 
4p_6^2=-4m_{out}^2=[2p_1-2p_3]^2
=4(-m_{in}^2-m_{out}^2-2m_{out}m_{in}\hat{p}_1\cdot\hat{p}_3)\nonumber
\eqe 
by mass conservation we have that $m_{in}=2m_{out}$, so, replacing it in the above equation we end up with 
$$\hat{p}_1\cdot\hat{p}_3+1=0$$
which is again the localization condition to $\hp_1\to \hp_3$.

\begin{figure}
\begin{center}
\includegraphics[scale=0.5]{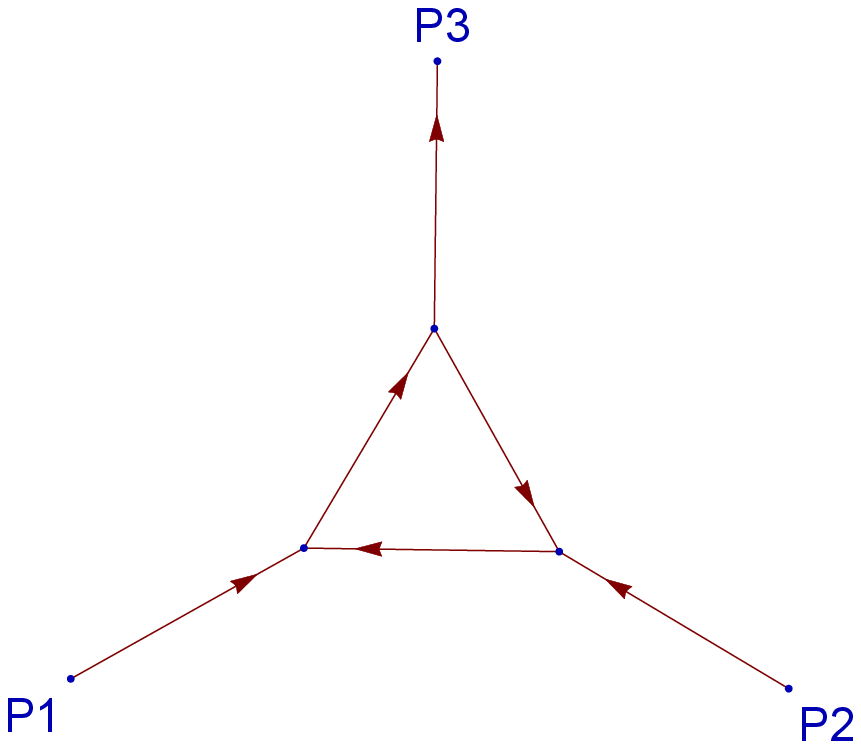}\hspace{3cm}\includegraphics[scale=0.45]{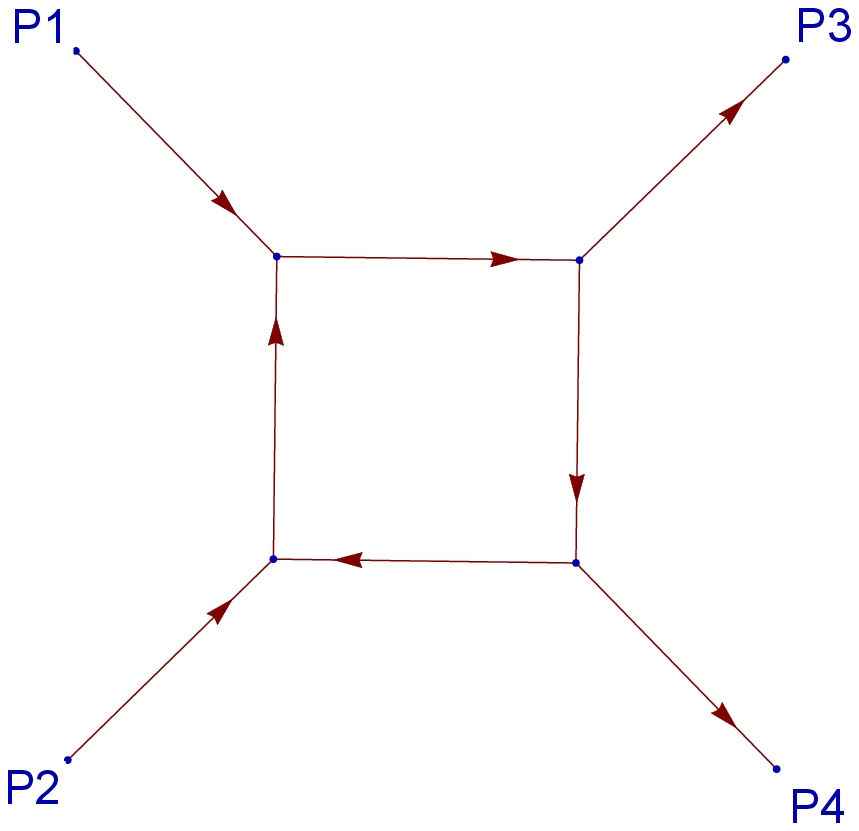}
\caption{One-shell one-loop diagrams}
\label{}
\end{center}
\end{figure}

\vspace{5mm}

{\it One-loop on-shell graphs}

Finally, let us consider the one-loop on-shell triangle and box graphs displayed in figure 5.
For the triangle, after imposing mass-conservation on every vertex, we can write the amplitude as (notice that $m_{out}=2m_{in}$)
\eqa
 A(p_i)=\int dp_{\ell}\delta\left(p_{\ell}^2+m_{\ell}^2\right)\delta\left((p_1+p_{\ell})^2+(m_{in}+m_{\ell})^2\right)\delta\left((p_2-p_{\ell})^2+(m_{\ell}-m_{in})^2\right)\nonumber
\eqae 
The first delta function constraint the loop momentum to be on-shell, and hence the integration  $dp_{\ell}$ above corresponds to a new $AdS_3$ point. After some algebra we can see that the second delta and third delta functions enforce the localization of $\hp_1\to \hp_{\ell}$ and $\hp_2\to \hp_{\ell}$ respectively, and momentum conservation fix $p_3=p_1+p_2=2m_{in} \,\hp_{\ell}$.
 
The remaining condition from momentum conservation,
\eq 
p_3^2=-4m_{in}^2=(p_1+p_2)^2=-4m_{in}^2\,,
\eqe
it is satisfied by mass conservation and therefore does not impose any additional constraint. 

The box is slightly more involved, but the argument applies similarly. The on-shell loop momenta condition and vertex mass conservation (taking all the external masses to be the same magnitude), allow us to write the amplitude as,
\eqa
 A(p_i)&=&\int dp_{\ell}\delta\left(p_{\ell}^2+m_{\ell}^2\right)\delta\left((p_1+p_{\ell})^2+(m_{in}+m_{\ell})^2\right)\times\nonumber\\
&&\delta\left((p_1+p_{\ell}-p_3)^2-m_{\ell}^2\right) \delta\left((p_2-p_{\ell})^2+(m_{\ell}-m_{in})^2\right)\,,
\eqae
as before, the first delta function forces the loop momenta to be on-shell, the second on the first line and the second on the second line, enforces $\hp_1\to\hp_{\ell}$ and $\hp_2\to \hp_{\ell}$ respectively. The first one on the second line can be carried out similarly leading us to,
\eqa
\frac{1}{(2m_{in}^2+2m_{in}m_{\ell})}\delta\left(1+\hp_{\ell}\cdot\hp_3\right)\,,
\eqae
hence implying $\hp_3\to\hp_{\ell}$. The remaining momentum conservation delta function reads then,
\eq 
p_4^2=-m_{out}^2=(p_1+p_2-p_3)^2=p_{1}^2=-m_{in}^2\,,
\eqe
which is trivially satisfied by mass conservation. Notice that we can extend these procedure for general polygons at one-loop, i.e. beyond four-points unlike as for unitarity cuts, since the delta functions are constraining the external data instead of the loop-momentum.

In summary, imposing the mass conservation condition coming from momentum conservation in five dimensions, plus momentum conservation in 4d lead us to a Witten contact term in $AdS_3$ for the massive exchanges. At this point, is worth to make a couple of remarks. Contact Witten diagrams are also known as $D-$functions, as has been defined in \cite{DHoker:1999kzh}, and they can be elegantly  representated in terms of Mellin amplitudes \cite{Penedones:2010ue}. So far we have considered all incoming particles to be adjacent and similarly for all the outgoing particles. This has been done in order to guarantee that we always have massive propagators, in the case where there are alternate incoming and outgoing particles, it is possible to produce massless propagators due to mass conservation. This case deserves special attention and is going to be discussed later on this note.

\section{A 2$d$-2$d$ duality}
We see that in the above construction, mass conservation plays an essential role. While this appears unnatural in a four-dimensional point of view, it arises naturally if we consider it as a five-dimensional massless $\phi^3$ theory, where the fifth momenta can be identified as the four-dimensional mass: $\vec{p}^{(5)}=(p, m)$. Momentum conservation in fifth dimension will then ensure that all masses are conserved at each vertex. At this point, this appears to be mere cosmetics. However, by reinterpreting the kinematics as massless, we can utilise the Cachazo, He, and Yuan (CHY) representation to reproduce the factorization singularity.

Recall that for $\phi^3$ theory, we can reconstruct its S-matrix by integrating over the moduli space of punctured points on the Riemann sphere~\cite{Cachazo:2013iea}, where the integrand is given by double cycles 
\eq\label{phi3chy}
m^{\phi^3}(\alpha,\beta)=\int\prod_{a=1}^n\frac{d\sigma_a}{{\rm SL}(2,\mathbb{C})}\,\frac{1}{(\sigma_{\alpha_1,\alpha_2}\cdots\sigma_{\alpha_n,\alpha_1})(\sigma_{\beta_1,\beta_2}\cdots\sigma_{\beta_n,\beta_1})}\prod_{i}\delta\left(\sum_{j\neq i}\frac{s_{ij}}{\sigma_{ij}}\right)\,,
\eqe
where $(\alpha,\beta)$ are any couple of permutations over the set of labels $\{1,2,\cdots,n\}$, $\sigma_i$ are the positions of the punctures on the sphere, $\sigma_{ij}=\sigma_i-\sigma_j$ and $s_{ij}=(p_i+p_j)^2$. 
By relabelling the puncture locations on the sphere as,
\eq
\sigma_a=\frac{\tau}{u_a},\quad a\in L,\quad \quad \sigma_a=\frac{v_a}{\tau},\quad a\in R,
\eqe
where $L$ and $R$ define the subset of index $L=\{1,\cdots,n_L\}$ and $R=\{n_L+1,\cdots,n\}$, it can be shown that the product of scattering equation delta functions factorises and the integrand in \eqref{phi3chy} takes the form:
\eq\label{factorchy}
-(u_{1,2}u_1u_2v_{n-1,n}v_{n-1}v_{n})^2\frac{d\tau^2}{\tau^2}\,
m^{\phi^3}(\alpha_L,L\,|\beta_L,L)(\{u\})\,\delta\left(p_R^2-{\over}\tau^2 F\right)\,m^{\phi^3}(\alpha_R,R\,|\beta_R,R)(\{v\})\,.
\eqe
Here $F$ is some $\tau$ independent polynomial, and 
\eqa
m^{\phi^3}(\alpha_L,L\,|\beta_L,L)(\{u\})&=&\int\prod_{a=3}^{n_L}\frac{du_a}{{\rm SL}(2,\mathbb{C})}\,\frac{1}{(u_{1}u_{1,\alpha_L(2)}\cdots u_{\alpha_L(n_L-1),\alpha_L(n_L)}u_{\alpha_L(n_L)})}\times\nonumber\\
&&\frac{1}{(u_{1}u_{1,\beta_L(2)}\cdots u_{\beta_L(n_L-1),\beta_L(n_L)}u_{\beta_L(n_L)})}\prod_{a\in L\setminus\{1,2\}}\delta(f^a_L)
\eqae
where $(\alpha_L,\beta_L)$ are any couple of permutations over the set of labels $L\setminus\{1\}$ and
 $f^L_a$ are the scattering equations for the subset of punctures in $L$,
\eq
f^L_a=\sum_{b\in L\setminus a}\frac{s_{a,b}}{u_{a,b}}\,.
\eqe
with a similar expression for $m^{\phi^3}(\alpha_R,R\,|\beta_R,R)(\{v\})$ but for the subset of elements $R$ and variables $v_a$. 

On the factorization kinematics $p_R^2=0$, the delta function $\delta\left(p_R^2-{\over}\tau^2 F\right)$ localises $\tau^2$ to vanish and hence correspond to the pinch limit of the Riemann sphere. Now we can easily turn this around and consider the pinch limit directly in the CHY integrand, which via the scattering equations will enforce the kinematics be in the factorization limit. For example, at four-points, one would have:
\eq
\delta(s_{12})=\int \frac{d\tau^2}{\tau^2}\delta\left(p_R^2-{\over}\tau^2 F\right)\tau\delta(\tau)\,.
\eqe
At higher points, one can reproduce all factorization singularities by successively pinching the Riemann sphere, i.e. for $n$-factorization singularities one simply performs a change of variable to make  $n$-pinch parameters manifest, and insert $n$-factors of delta functions that localise these $n$ parameters. The result of the integration would simply a product of $n$ factorization delta functions. 

Finally we parameterise the five dimensional momenta using our AdS coordinates, where there is freedom in the embedding of four-dimensions. Depending on the embedding, the internal moment can be massive or massless. Will will discuss the massless case in the next section. For the massive case, one finally establishes the following correspondence:
\eq
2d\;\;CFT\; \leftrightarrow  \prod_{i=1}^n\left(\int d^2z_i\frac{dy_i}{y_i^3}G_\Delta (y_i, z_i,\bar z_i; w,\bar w) \right)\; \int  d^{2n}\sigma_i \;{\rm CHY} \;\left(\prod_j \tau_j\delta(\tau_j)\right)\,.
\eqe 
where $\tau_j$ are the moduli of the Riemann sphere whose zero limit correspond to pinched limit.

\section{Massless singularities}
In reducing the five-dimensional representation to four-dimensions, one also naturally recover cases where the propagators are massless. For massless singularities the AdS points are no-longer localised to a single point. Rather we recover configurations with multiple bulk points similar to the factorization Witten diagrams. We will demonstrate this with the previous four-point example.

\begin{figure}
\begin{center}
\includegraphics[scale=0.6]{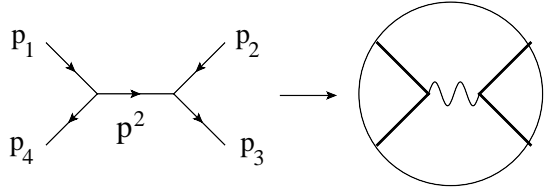}
\caption{The factorization singularity of $u$-channel diagram leads to a factorization Witten-like diagram in AdS$_3$ }
\label{fig2}
\end{center}
\end{figure}

Consider again the kinematics of the external lines in fig\ref{fig1} but with factorization in the $t$-channel. Now mass conservation implies that the internal propagator is massless, and hence 
\eq
\delta((\hat{p}_1-\hat{p}_4)^2)=\frac{1}{2}\delta(\hat{p}_1\cdot\hat{p}_4+1)
\eqe
The factorization singularity now identifies $\hat{p}_1$ to $\hat{p}_4$ and $\hat{p}_2$ to $\hat{p}_3$. However as momentum conservation is simply $\hat{p}_1+\hat{p}_2-\hat{p}_3-\hat{p}_4=0$, it is trivially satisfied in this limit, and hence one ends up with two free points in AdS$_3$, i.e. putting this together we got 
\begin{align}
\mathcal{ \tilde A}_4(w_i,\bar w_i)\sim \lambda^2\frac{\epsilon^{1/2}}{m^{4}} \int_0^\infty   {dy_1 \over y_1^3} \int dz_1^2 \int_0^\infty   {dy_3 \over y_3^3} \int dz_3^2 \prod_{i=1}^4 G_{\Delta_i} (y_i , z_i,\bar z_i ;  w_i,\bar w_i)
\,\delta(0)\tilde{A}(p)\,,
\end{align}
where $\tilde{A}(p)$ represents the residue of the flat space amplitude around the singularity in the $t-$ channel, whose form now depends on the spin of the internal particle. This expression is a factorization Witten-like diagram, as shown in fig.(\ref{fig2}). 
\vspace{0.5cm} 

The residue of the massless singularity is now sensitive to the spin of the exchanged particle. For spin-0 the residue is simply 1, while for spin-1 one has $\hat{p}_3\cdot\hat{p}_1$. These do not appear to correspond to any bulk to bulk propagators in AdS which are non-rational functions of $\hat{p}_3\cdot\hat{p}_1$, and thus cannot correspond to any single operator exchange. We will further comment on this in the conclusions.

\section{Conclusions and outlook}
In this note, we have considered the correspondence between four-dimensional S-matrix singularities and two dimensional CFT. This is in the context of the Pasterski, Shao and Strominger construction, where one replaces the usual S-matrix plane waves with conformal primary wave function. The kinematic space is conveniently parametrized by AdS$_3$ coordinates, and thus correspond to $n$- AdS bulk points. The focus on the factorization singularities is then the first step to understand the implications of the dynamic properties of the S-matrix.  

We show that for massive scalar theories, if mass conservation is implemented at each cubic vertex, the massive factorization singularities along with over all momentum conservation, will localizes the $n$-copy of AdS bulk points to a single point thus forming a contact Witten diagram. Note that in this case, the different spin exchanges simply degenerates to an overall normalisation constant. For massless singularities, the result is a factorization Witten diagram with polynomial residues whose degree depends on the spin of the exchange particle, which are rational polynomials. However, these do not correspond to the usual bulk to bulk massless spin-1 exchanges which would require non-rational functions of $\hat{p}_i\cdot \hat{p}_j$. One possibility is that the dimension of the exchange operator is continuous, and one is required to perform an integral along some contour for the dimensions. Another would be that this contributions no longer corresponds to single operator exchange. Finally, in this note we only consider massive scalars. For massive higher-spin amplitudes, the S-matrix will transform covariantly under the massive SU(2) little group. The meaning of this SU(2) on the CFT is not clear to us, and we leave these interesting questions to future studies.  

The fact that mass must be conserved at each vertex indicates that the constraints are more naturally embedded in massless constraint of five-dimensional kinematics, where the conservation is simply the extra dimensional Poincare symmetry. This identification also allows us to use the CHY representation for massless S-matrix, and reinterpret the factorization singularities as pinch limits of the Reimann sphere. This establishes an interesting 2d/2d correspondence.

\section{Acknowledgements}

We would like to thank Shu-Heng Shao and Ellis Ye Yuan for discussions. C.C . would like to acknowledge the Mainz Institute for Theoretical Physics (MITP) for its hospitality and support during the completion of this work. Y-t Huang would also like to thank the Institute for Advanced Study for its hospitality. Y-t Huang is supported by MOST under the grant No. 103-2112-M-002-025-MY3. The work of C.C. is supported in part by the National Center for Theoretical Science (NCTS), Taiwan, Republic of China.
\vspace{-0.3cm}

\bibliographystyle{JHEP}
\bibliography{CFDBib}
\end{document}